# σ/π character of H-bonding in water clusters

Aleksandr S. Bedniakov and Yulia V. Novakovskaya*

Moscow State University, Leninskie Gory 3/1, Moscow, 119991 Russia

E-mail: *jvn@phys.chem.msu.ru; jvnovakovskaya@gmail.com*

**Abstract:** Hydrogen bonds are typically treated as sufficiently localized directional intermolecular bonds, in which dispersion and electrostatic contributions can be distinguished. However, being formed chiefly due to the overlapping of p orbitals of electronegative atoms, the corresponding electronic bonds are characterized by both σ- and π-kind binding, the former determining the directionality of bonds, while the latter, the coupling of molecules and the collective effects in H-bond networks. The latter contribution was never considered previously and is predetermined by overlapping pre-lone pair orbitals of oxygen atoms. This is manifested in the peculiarities of the electron density distribution, which are quantified based on the analysis of magnetic shielding tensors of oxygen and bridge hydrogen nuclei and illustrated by the shapes of cluster orbitals of water aggregates.

## INTRODUCTION

The concept of hydrogen bond in the form accepted and conventionally used by the scientific community was formulated by L. Pauling [1]. It reflected the progress in understanding the character of the underlying effect that was achieved in nearly a quarter of a century. The key steps made by that time were as follows. The beginning of the discussion was due to T.S. Moor and T.F. Winmill [2] who supposed the possibility of the existence of weak unions formed by molecules such as water or ammonia due to the special kind of intermolecular bonds. Nearly a decade later in the works of M.L. Huggins [3] and W.M. Latimer and W.H. Rodebush [4], already the notion of hydrogen bonds appeared.

Nowadays, we can state that there are numerous (or even innumerous) compounds already discovered and synthesized or yet unknown and still to be found that involve functional groups, which can form hydrogen bonds and, in this way, gain particular characteristics inevitably manifested in various ways. All such groups involve an electronegative element (oxygen, fluorine, nitrogen) directly bound to a hydrogen atom (-OH, FH, =NH) and can be coordinated to allied groups. In addition to these so-to-speak classical variants, there are less typical examples when such groups are coordinated to either rare gas atoms, e.g., FH…Rg [5] or to π-electronic systems, particularly conjugated, e.g., X–H...π bonds (where X = O, N, C) [6]. However, in all the situations the counteragent can provide electron density sufficient for either stronger or weaker binding.

All the variants were summed up in the analysis of the milestone works in the field (see refs. in [7]); and the Task Group on Categorizing Hydrogen Bonding and Other Intermolecular Interactions has recommended the following definition of H-bonding [8]: "The hydrogen bond is an attractive interaction between a hydrogen atom from a molecule or a molecular fragment X–H in which X is more electronegative than H, and an atom or a group of atoms in the same or a different molecule, in which there is evidence of bond formation". X–H is the bond donor, while Y or Y–Z is the acceptor with an electron rich region, such as a lone pair of Y or a π-bonded electron pair of Y–Z.

The evidences of bond formation are collected in the lists of criteria and characteristics of X–H...Y–Z hydrogen bonds the definition is supplemented with. We would like to draw attention to the following [8]:

(i) the nature of forces that determine the formation of a hydrogen bond is chiefly electrostatic; they arise due to the charge transfer between the donor and acceptor and involve the dispersion contribution;

(ii) X–H covalent bond is polarized upon the H-bond formation, and the strength of the resulting H...Y bond increases with an increase in the electronegativity of X;

(iii) X–H...Y–Z hydrogen bond is characterized by NMR signatures such as the deshielding of proton in X–H, spin–spin coupling between X and Y, and the nuclear Overhauser enhancements;

(iv) H-bond energy correlates well with the extent of charge transfer between the donor and the acceptor.

There are other well-known features of hydrogen bonds, such as the elongation of X–H covalent bond and the relative shortening of H…Y intermolecular contact, which becomes more pronounced as the binding is strengthened; the decrease in the frequency of X–H stretching vibration accompanied by the appearance of additional vibrations; and so on. We selected the above listed criteria because all of them are related to the peculiarities of the electron density distribution in the particles, which are susceptible to H-bond formation, and the resulting charge redistribution caused by the H-bonding. And definitely it is the peculiarities in the electron density distribution in H-bonded clusters that are responsible for the so-called collective effects typical of H-bond networks. However, the above list as well as the general analysis of hydrogen bonding did not provide any clear physical explanation of the nature of the collectivity in H-bonded ensembles. In this paper we suggest an approach that sheds light on the problem.

Obviously, with an intention to clarify the physical grounds of some effect it is expedient to select objects that are most spectacular. The substances that can conditionally be named basic in view of the selectively involving the main H-bonding groups solely are hydrogen fluoride, ammonia, and water. Hydrogen fluoride molecule can act as a donor (*d*) of only one proton and an acceptor (*a*) of one or two protons in H-bonds. In some structures, bifurcated H-bonds can be formed when one proton is involved in two bonds, but such configurations are not as stable. Thus, extended planar or stepwise rings composed of hydrogen fluoride molecules, which act as *da* or *daa* coordination joints, are those structure elements that can appear most easily. In the case of ammonia molecule, the local coordination is nearly inversed, namely, the molecule can act as a donor of all three protons and an acceptor of only one proton in H-bonds (*da*, *dda*, or *ddda* coordination). And it is only water molecule that can act as a double donor and double acceptor of H-bond protons to provide the ultimate *ddaa* coordination, which is symmetrically balanced in view of the H-donation and withdrawing. This symmetry predetermines the existence of a branched hydrogen-bond network with quite stable tetrahedral coordination joints. Naturally, as prevailing this variant is observed in ices, while in a liquid phase thermal dynamic perturbations inevitably cause local changes in coordination numbers to three (*dda* or *daa*) or even two (*da*), but still the electrostatic potential of a water molecule determined by two OH bonds and a lone electron pair of oxygen always results in the spatial correlation of molecular orientations typically referred to as short-range order.

Taking into account the electron density fraction that can be spent on the formation of intermolecular H-bonds and the numbers of bonds that can be formed by a molecule as a donor and acceptor of protons, one can see that water molecules represent the most interesting example. And it is water clusters that can provide the necessary and sufficient information about the character of the electron density redistribution that underlies the H-bond formation and the correlations in molecular ensembles.

**METHODICAL**

To solve the problem formulated, it is necessary to clarify peculiarities of the electron density redistribution that accompanies the formation of hydrogen bonds. In the basis of natural orbitals, the total electron density of a molecular system $\rho(r)$ is given by a sum of the weighted densities that correspond to individual $\phi_i(\mathbf{r})$ orbitals:

$$\rho(\mathbf{r}) = \sum_{i=1}^{P} n_i \, |\phi_i(\mathbf{r})|^2, \qquad (1)$$

where $n_i$ weights are the effective occupancies of the orbitals. In the case of HF (Hartree-Fock) or KS (Kohn-Sham, or density functional theory, DFT) approaches, $P$ is equal to the number of electrons ($N$), and all $n_i = 1$ for $i = 1 \ldots N$ (or $n_i = 2$ and $P = N/2$ in the restricted variants when $\phi_i(\mathbf{r})$ orbitals coincide in pairs), however, the principal difference between the methods is the possibility of interpreting all the occupied HF orbitals as the amplitude probabilities of the electronic charge distributions, which cannot be proved for the solutions of KS equations in the DFT approach. A similar problem is faced in the case of any multiconfigurational method when the closeness of $n_i$ values to one or two does not give a strict basis rather than a believe that the corresponding $|\phi_i(\mathbf{r})|^2$ functions can be treated as reliable approximations of the actual charge distributions of individual electrons. Nevertheless, experience shows that such an interpretation is quite reasonable at least in the situations when the total electron density distribution is not characterized by some so-to-speak exotic features, which is the case of most of the compounds that involve chiefly atoms of the second-row elements (such as fluorine, oxygen, nitrogen, carbon) and are able to form hydrogen bonds. Therefore, we are going to adhere to such a hypothesis, the more so that for molecular water clusters, the HF orbitals (their approximations by linear combinations of atomic orbitals) are very close to the natural orbitals constructed at different post-HF levels whose occupancies are larger than 0.03, or, in other words, close to 2. The latter is predetermined by the essentially correct representation of the total electron density distribution in one-determinant approximation (except for the inevitable inherent problems of the method that are not crucial for the aspects touched in this paper).

The phenomenological analysis of the data obtained is carried out within the following paradigm. The formation of hydrogen bonds is accompanied by numerous manifestations in the electronic-nuclear structure and energetic characteristics that are very substantial and sometimes even approach those of weak covalent bonds. This fact distinguishes hydrogen bonds among all the residual intermolecular contacts and shows that perturbation theory applicable in most other situations is not as suitable for the description of hydrogen bonds. The preferable and more founded variant is the variational approach. It implies that the quantity of interest can be approximated by a linear combination of some basis functions. When the object is a hydrogen bond, which is formed between the originally independent particles and is electronic in nature, a reasonable choice for the basis functions is the set of molecular orbitals of the individual particles involved in the formation of the bond orbitals. Then, the following scheme akin to a typical MO LCAO variant can be used for the phenomenological description and analysis.

Orbitals that represent the electronic charge distribution in the resulting molecular cluster (composed of individual particles), which are referred to as cluster orbitals (CO) below, can be approximated by linear combinations of molecular orbitals (LCMO) of the cluster constituents. In the figures below, cluster orbitals are drawn with a contour threshold value of 0.01 a.u. To give a tentative idea about the relative energies of the corresponding electronic charge distributions, their HF orbital energies are used.

These cluster orbitals are no more than approximate functions the summed squared absolute values of which reasonably approximate the total electron density of the cluster and which, at the same time, enable us to judge the overlapping character of the molecular orbitals. This aspect is important in view of the classification of the resulting one-electron functions as resembling conventional σ or π orbitals. However, such information is definitely insufficient for formulating unambiguous conclusions about the general character of the electron density distribution in the clusters and characterizing peculiarities of the density redistribution that reflect the character of H-bonding as electronic bonding. Such conclusions need more reasonable grounds, which can be provided by the estimated response of the molecular system in question to an external perturbation where it is the electron density distribution that gives rise to the effect.

The most reliable characterization of the charge distribution peculiarities is the one that reflects the energy change provided by the interaction of a particular nucleus with the local magnetic field, the strength of which depends on the nucleus shielding produced by the electron density distribution around it:

$$\Delta E = -(\mathbf{\mu}_J, \mathbf{H}_J),$$

where $\mathbf{\mu}_J$ is the magnetic dipole moment of atom J, $\mathbf{H}_J = (\mathbf{1} - \mathbf{\sigma}_J)\mathbf{H}$ is the local magnetic field at the J nucleus; and $\mathbf{\sigma}_J$ is the shielding tensor, the trace of which enables one to estimate the isotropic shielding:

$$\sigma_{J,iso} = \frac{1}{3} tr(\mathbf{\sigma}_J),$$

while the principal components ($\sigma_{\alpha\alpha} \geq \sigma_{\beta\beta} \geq \sigma_{\gamma\gamma}$), to judge the anisotropy of shielding:

$$\Delta\sigma_J = \sigma_{\alpha\alpha} - \frac{1}{2}(\sigma_{\beta\beta} + \sigma_{\gamma\gamma}).$$

The responses of both $^{17}O$ and $^{1}H$ nuclei to the external magnetic field can provide information about the hydrogen bonding, since the shielding of all the nuclei involved in the bonds should change upon the bond formation. However, as one could see, only the deshielding of bridge protons is mentioned in the characteristics of hydrogen bonds. At the same time, in practical applications, $^{17}O$ NMR is more informative (compared to $^{1}H$ NMR) primarily due to the non-exchangeable nature of O sites and the broad range of shielding parameters [9,10]. For example, $^{17}$O NMR spectroscopy was used for the investigation of H-bonding in water, proteins, biological macromolecules, and aqueous solutions [11-14]. Furthermore, it was used for assessing the dynamics of bound water in salt solution [15] and the state of water molecules in solid hydrates [16, 17]. Moreover, it is water that is used as a reference compound for estimating the chemical shifts of $^{17}$O in diverse substances. Therefore, more detailed information about the shielding of oxygen nuclei in water molecules depending on the degree of their involvement in H-bond networks is necessary.

However, in the case of individual water clusters, such an experiment in a gas phase seems impossible at the present moment. Even if it were possible, of all the shielding tensor information, only the orientation-averaged isotropic shielding can be measured experimentally. But it provides a gross characteristic of the charge distribution with no sufficient information about the local electron density redistribution, which is reflected in the anisotropy, or more generally in the principal values of the tensors. Then, simulations of water clusters are the sole source of the information required. Moreover, the simulations should provide sufficiently accurate estimates, which is possible only when the tensor elements are found based on the analytic expressions for the energy derivatives with respect to the spin and magnetic field components with the use of the gauge-independent atomic orbitals technique [18, 19]. It is this approach we used in our investigation; and analyzed the atomic shielding tensors paying attention to (i) the absolute and relative values of the principal values of the tensors; (ii) the changes in the values upon the involvement of a molecule in different numbers of hydrogen bonds; and (iii) the directions of the tensor eigenvectors and their correlation to the directions of covalent and hydrogen bonds in the systems in question.

Starting with the pioneering works in the field, the change of the isotropic shielding of oxygen nuclei in water with respect to that in the gas phase was determined as -36 ppm, being decreased (supposedly proportionally) with an increase in temperature to about -26 ppm at 215°C [20, 21]. According to the revised experimental shielding scale for oxygen, the isotropic shielding equals 287.5 ppm in liquid water and 323.6 ppm in gas phase [22]. Note that it is partial breakage of hydrogen bonds that takes place with an increase in temperature and the drastic destruction of the H-bond network during evaporation. Then, the gradual aggregation of water molecules accompanied by the formation of progressively more branched and extended three-dimensional H-bond network should be reflected in a gradual change in the shielding parameters of the nuclei depending on the local coordination of water molecules in the resulting aggregates.

Previous calculations of the shielding parameters for water molecules and their clusters were carried out at different levels, including Hartree-Fock (HF) and density functional (DFT) approximations with different (local, generalized gradient and hybrid exchange-correlation functionals), in combination with the gauge including projector augmented wave (GIPAW) pseudopotential method or the polarizable continuum model (PCM), as well as the second order of the Møller-Plesset perturbation theory (MP2) with various basis sets, in comparison to the restricted active space self-consistent field calculations in combination with rovibrational corrections [23-26]. It was found that continuum models cannot provide any reliable estimates of the actual shielding parameters. In the case of cluster approaches, the role of polarization functions was stressed. The absolute values almost always differed from the aforementioned recommended ones except for the situations when dynamic (e.g., rovibrational correction) effects were taken into account; and the typical average deviation is about 30 ppm for $^{17}$O nuclei.

In what concerns the parameters obtained, either the full shielding tensors were considered, but for individual molecules with a certain averaging for the molecules in a liquid phase [25], or only isotropic shieldings were discussed for the molecules with different local coordination neighborhoods involved in diverse structure elements (rings) [26]. Unfortunately, a systematic analysis of the shielding tensor parameters for the water molecules with various coordination neighborhoods is absent. And it is only this analysis that can enable one to judge the character of the electron density redistribution depending on the character of donor-acceptor interactions between the molecules.

Taking into account the literature data with the aim to obtain quantitative information that can be used later as a reference for describing large H-bonded ensembles of water molecules and foreign particles, either relatively small ionic or relatively large biological, we selected the following level of modeling. All nonempirical simulations were carried out at the MP2 level with the 6-31+G(d) basis set on oxygen and 6-311+G(p) set on hydrogen atoms. The basis is simultaneously sufficiently flexible to provide a reasonable description of the electron density redistribution upon the formation of both covalent and hydrogen bonds and sufficiently compact to avoid the linear dependence for the larger systems. It is worth noting that the approach can by no means provide the most accurate energy estimates and the best possible approximation of the total electron density distribution. However, it was intentionally selected as a compromise, which, on one hand, provides reliable approximations of the one-electron orbitals and reliable changes in the approximations depending on the local neighborhoods of the molecules, and, on the other hand, a clear interpretation of the underlying effects (when the relative increments of different atomic sites can easily be related to the σ or π character of the corresponding electronic binding). Thus, being not deprived of systematic errors, it gives reasonable relative values, which form the basis for the methodological analysis we carry out.

All the configurations considered correspond to the energy minima, which is confirmed by the normal-coordinate analysis. All calculations were carried out with the use of Firefly QC package [27], which is partially based on the GAMESS (US) [28] source code, and CFOUR program package [29, 30] and visualized with Chemcraft software [31].

Taking into account that water molecules are involved in innumerous kinds of cluster systems discussed in innumerous papers, it is absolutely unreasonable and impossible to quote all of the works in the field. At the same time, it is disrespectful to quote only some of them and neglect the others. Therefore, except for the above quotation of the works devoted to the shielding tensor calculations, the discussion below contains no reference to independent studies. A representative list of references can be found in the aforementioned technical report of the IUPAC Task Group [7].

**RESULTS AND DISCUSSION**

*Character of cluster orbitals*

Tentative conclusions about the character of the electron density redistribution upon the formation of H-bonded clusters of water molecules can be drawn from the analysis of cluster orbitals. In terms of the CO LCMO approach, one can distinguish variants of MO overlapping akin to typical $\pi$ and $\sigma$ overlapping of atomic orbitals in the conventional MO LCAO scheme. Possible variants of bonding and characters of cluster orbitals are illustrated below by examples of water clusters with diverse spatial organization. Not of all them correspond to the global adiabatic potential minimum; and these were selected based on the following ideas.

In any three-dimensional H-bond network of water molecules, one can distinguish rings. The most frequently met are tetra-, penta-, and hexamolecular rings. The rings can be connected via a common tetracoordinate molecule or via a common edge, which is actually a common hydrogen-bond O–H…O sequence of atoms. Additionally, all stable individual rings (that correspond to the global potential energy minima) are characterized by a directional sequence of hydrogen bonds of (H)O–H…(H)O–H… kind. It is this kind of rings named homodromic that was selected as basic for constructing model structures in our study. The structures are shown in Figure 1. These are (i) plain rings; (ii) rings joined via a common tetracoordinate molecule; (iii) rings joined via common bimolecular edges in slightly convex-concave structures; and (iv) 3D cages composed of fused rings in more or less regular variant.

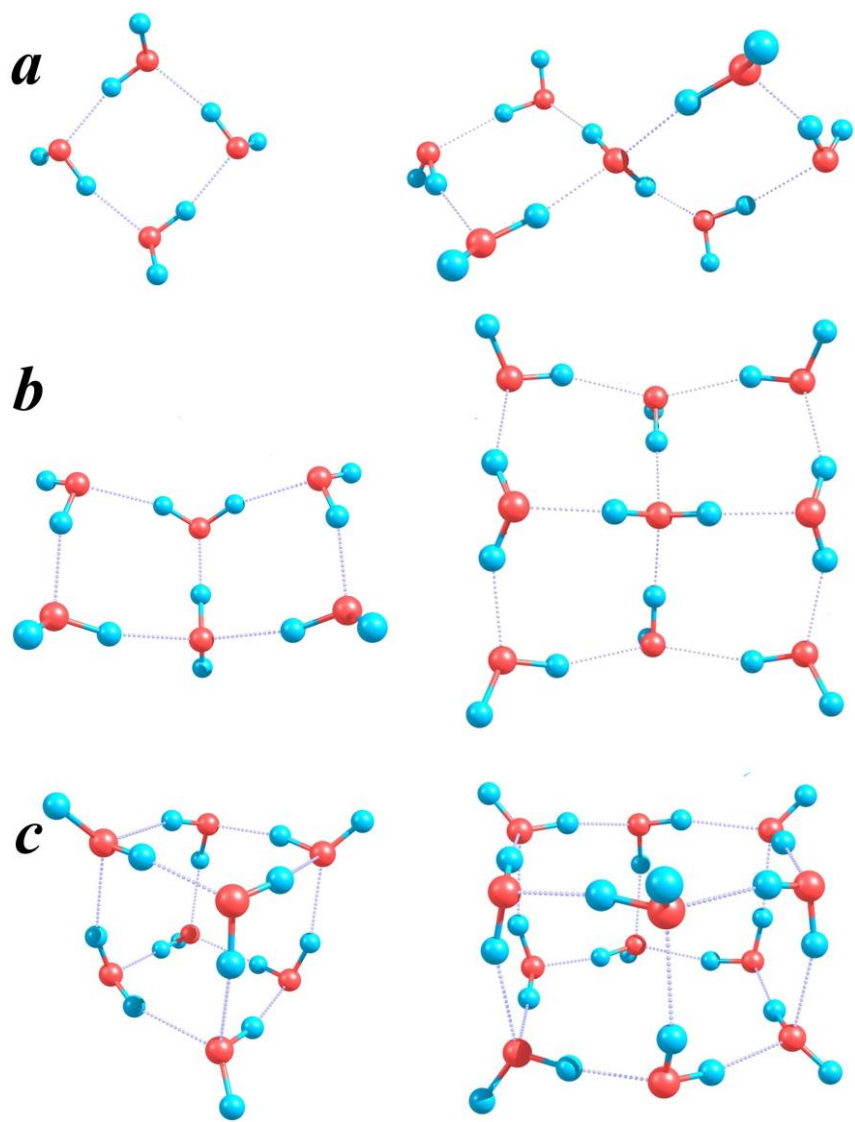

**Figure 1.** Cluster structures: (a) plain-ring tetramer and heptamer composed of vertex-joined tetrameric rings; (b) convex-concaved bicyclo hexamer and tetracyclo nonamer composed of fused rings; and (c) octamer and regular dodecamer cages composed of fused homodromic tetrameric and hexameric rings.

Let us start with a plain ring as characterized by the least noticeable effect produced by the overall non-planarity. If one takes a look at the orbitals of an individual tetramer (Figure 2), indeed two kinds of orbitals can be found. In one variant, MOs that can roughly be approximated either by $2p_x(O) + (1s(H^1) - 1s(H^2))$ linear combination (if xOz is the nuclear plane of the molecule, and Oz axis is its symmetry axis) or by $2p_z(O) + (1s(H^1)+1s(H^2))$ overlap; and these are σ-kind cluster orbitals with the energies of -0.731 and -0.620 a.u. respectively. In the other variant, MOs that are essentially $2p_y(O)$ lone-pair orbitals overlap according to π-kind (-0.503 a.u.). Not to overload the illustration but to give an idea of how the orbitals look like, Figure 2 shows cluster orbitals of the prevailing bonding character solely. As one can notice, orbitals of both kinds are quite similar to conventional intramolecular orbitals, but the corresponding charge densities are substantial (and constant-sign) within the regions of hydrogen bonds. The density close to the oxygen nuclei is noticeably higher than around H-bond protons or close to the bond midpoints, but generally, the orbital characters are quite conventional. Moreover, (i) the bonding contribution that comes from π-kind orbitals is comparable to that of σ ones and (ii) the bonding provided by both kinds of orbitals spans the whole ring. Note also that at a relatively large contour threshold value of 0.01 a.u., the π-kind orbital (Figure2, c) has smaller "gaps" (that disappear at a more tight threshold value of 0.005 a.u.), which reflects its stronger delocalization.

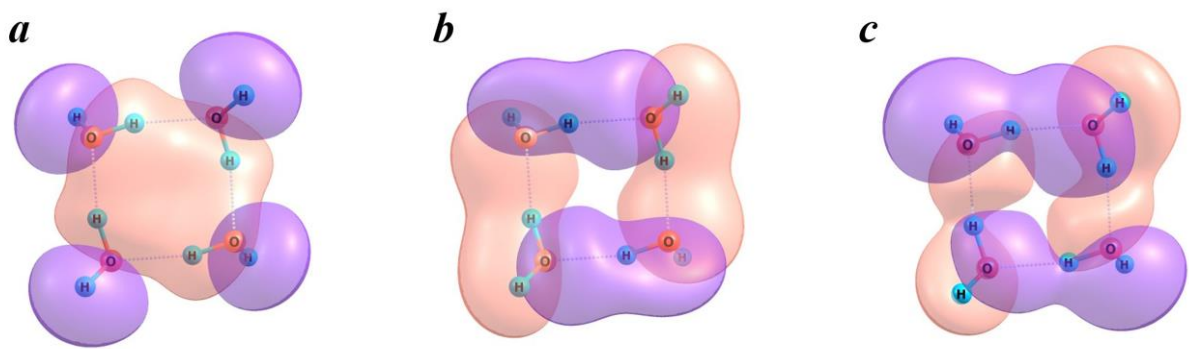

**Figure 2.** Cluster orbitals of (H$_2$O)$_4$ ring tetramer of (a, b) σ and (c) π kinds (the orbitals are shown as transparent to make their spatial character clearer).

The delocalization is akin the one typical of conjugated hydrocarbons where single and double bonds are alternated. In water tetramer considered, there is an alternation of intramolecular O–H covalent and intermolecular H…O hydrogen bonds. Undoubtedly, these cannot become comparable in length, but the relative elongation of the covalent bonds and shortening of the hydrogen-bond contacts upon the appearance of a closed sequence of alternated covalent and hydrogen bonds is quite clear. The covalent O–H$_b$ bonds (where b subscript denotes the bridge proton) are increased by 0.012 Å, while the O…H$_b$ distances are shortened by ca. 0.150 Å compared to a water dimer stabilized by a single H-bond.

At the same time, the lower electron density within the intermolecular contact regions and the inclination of the pre-lone pair orbitals of oxygen atoms (which mainly contribute here) with respect to the mean oxygen plane of the cluster because of the inclination of the individual molecular planes make the delocalization a bit different from that noticed in organic compounds. As follows from Figure 2, within each H-bond domain, either positive- or negative-sign segments of the atomic orbitals overlap to a stronger degree, so that the π-kind density distribution is no longer uniform as regards the oxygen plane. However, as follows from Figure 2, the total contribution from this kind of bonding to the total electron density within the intermolecular contacts is substantial.

In a heptamer composed of two joined tetramers with one common vertex molecule (Figure 1a), both tetrameric rings are similar to the individual homodromic tetramers, but the rings are mutually orthogonal. In this situation, the electronic charge distribution in the cluster is described by the orbitals, which look like combinations of the σ or π cluster orbitals of two tetrameric subunits (Figure 3). And in the case of π-orbitals, a smooth flexion between the two ring subunits can be noticed. The alternating sequence of covalent and hydrogen bonds is continuous; and the wavy changes in the charge distribution make the π-coupling between the ring subunits quite possible. The energies of the cluster orbitals shown in Figure 3 are -0.734, -0.708, and -0.626 a.u., which qualitatively illustrates quite an expected trend. Insofar as the rings are orthogonal, the orbitals that can be approximated by linear combinations of the molecular orbitals of the same kind can no longer be energetically favorable due to the appearance of zeros within the domain of the common molecule. Only some special combinations should be predominantly bonding. Therefore, the very narrow energy gap between the two kinds of the bonding σ-cluster orbitals in individual tetramer increases to 0.026 a.u. at a relative stabilization of the higher-energy one. At the same time, the energetic position of the bonding π-orbital becomes much better. Thus, despite such visually unfavorable geometrical configuration, the π-binding provides an additional stabilizing effect compared to an individual ring.

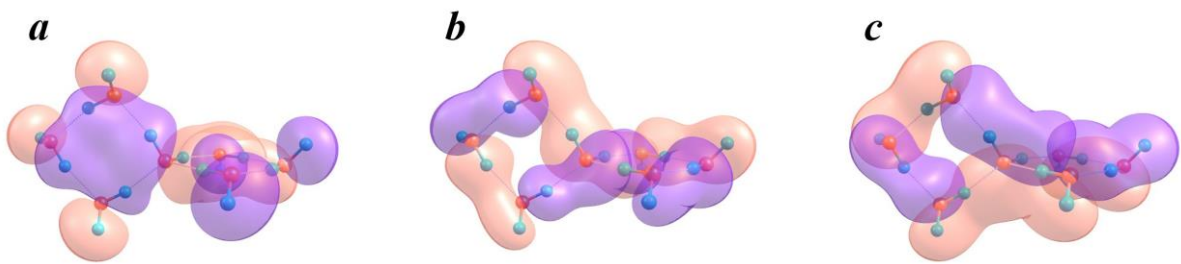

**Figure 3.** Cluster orbitals of $(H_2O)_7$ heptamer composed of two homodromic tetrameric rings with a common tetracoordinate molecule of (a, b) σ and (c) π kinds (the orbitals are shown as transparent to make their spatial character clearer).

Now, let us look at a hexamer composed of two fused homodromic tetramers with an H-bond common for the two rings (Figure 1b). At such joining of the rings, it is still possible to discriminate σ and π-kind cluster orbitals. Figure 4 shows those that are of the predominantly bonding character in the H-bond domains. All of them are definitely combinations of the orbitals of individual rings. They clearly envelop both rings despite the nonplanarity of the whole structure. Naturally, σ orbitals have lower energies (-0.737 and -0.712 a.u. for two cluster orbitals that can be viewed as superpositions of the orbitals that involve $2p_x$ and $2p_z$ AOs of oxygen atoms respectively (shown in Figure 2a, b) vs. -0.626 a.u. for the π cluster orbital), but it is π-orbitals that predetermine the spatial correlation of H-bonds, which embraces the whole hexamolecular structure. Note that compared to the individual tetramolecular ring, the energies of both σ and π orbitals decrease, which can be treated as a manifestation of the stabilizing effect provided by the delocalization over two similar structural units.

It is worth noting also that the above configuration has a total electronic energy 0.6 kcal/mol higher than that of the global minimum structure. The latter can also be viewed as two fused tetramolecular rings, but only one of them is homodromic, which corresponds to the local reorientation of two central molecules common for the two rings. The above favorable energy change (which can nevertheless be leveled off already at a room temperature due to the thermal perturbation) is due to the slightly less strained position of the molecule that acts as a double proton donor. The reorientation results in a partial loss of the continuous delocalization of the orbitals within the common O–H…O edge of the tetrameric rings, but the effect can still be traced in the here homodromic hexamolecular ring.

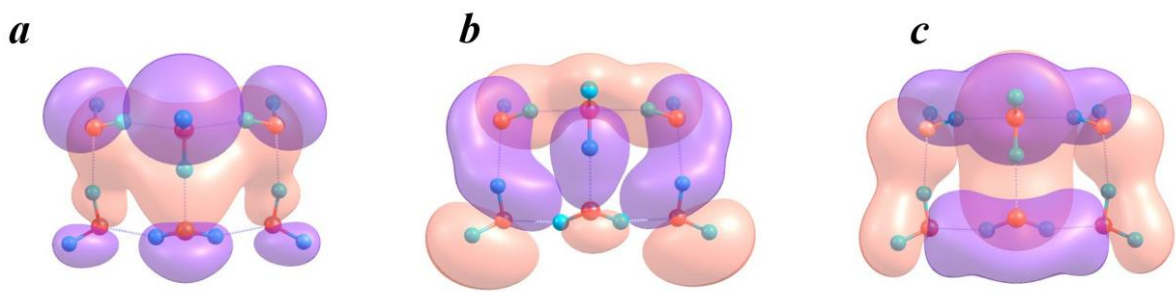

**Figure 4.** Cluster orbitals of $(H_2O)_6$ bicyclo-hexamer of (a, b) σ and (c) π kinds (the orbitals are shown as transparent to make their spatial character clearer).

When the number of rings is increased to four and they are annulated in such a way that there is one common molecule for all (Figure 1b), any ring should have two common edges with the two neighbors, and, hence, there should be four common O–H…O sequences, or more accurately, O–H…O…H–O and O…H–O–H…O crossing sequences. This is an untypical structure, which is rather an artificial construct; and we analyze its electronic structure only to illustrate which consequences follow from the annulation of homodromic rings when there are molecules involved in several such rings. Some cluster orbitals are shown in Figure 5. The cluster is convex-concaved with a saddle-like structure; and the O…O…O angles are 128 and 142 in one direction and 126 and 136 in the orthogonal. In the lower energy cluster orbitals of the same basic kinds as in the individual tetramer, the monomeric prototypes (or components) can easily be distinguished. However, now it is only the lower-energy one that is of purely σ-kind. The two residual orbitals shown in Figure 5 are of the prevailing σ- and π-kinds respectively only in the region of periphery molecules and of the complementary π- and σ-kinds respectively within the central trimolecular segment. The orbital energies are -0.728, -0.627, and -0.508 a.u. respectively, which shows (when compared to the orbital energies of the individual tetramer) that the opposite changes in those of σ-kind counterbalance each other, so that their total energetic contribution remains at nearly the same level, and that of π-kind has nearly the same energy. This means that such a complex wavy configuration is stabilized due to nearly the same effects as those in an individual ring. Even such an unfavorable annulation of homodromic rings does not noticeably distort the coupling within the rings, the smooth sewing of which is provided by the complementary variant of overlapping.

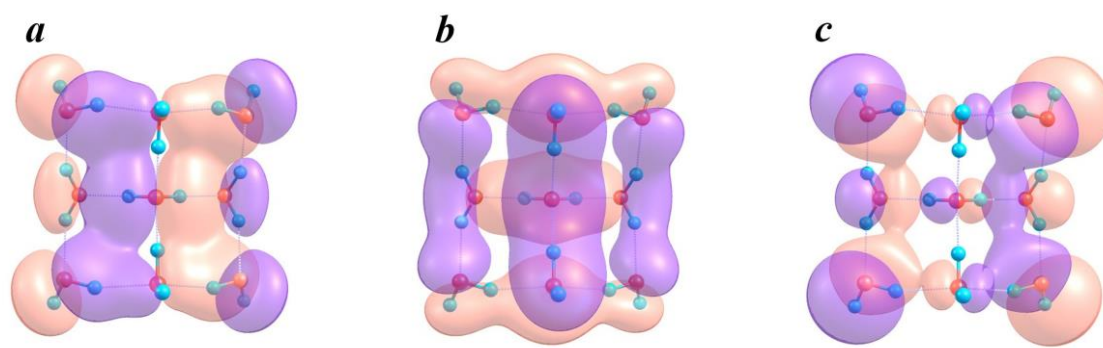

**Figure 5.** Cluster orbitals of $(H_2O)_9$ tetracyclo-nonamer of predominantly (a, b) σ and (c) π kinds (the orbitals are shown as transparent to make their spatial character clearer).

Now, let us turn to more physically reasonable three-dimensional closed cages composed of molecular rings; and start with the known stable octamer with a cubic conformation (of $S_4$ symmetry, Figure 1c). In the cube, two faces are tetramers with alternating covalent and hydrogen bonds, while all the residual necessarily involve two molecules, which are either double proton donors or double proton acceptors within particular rings. Then, it is reasonable to base the conditional classification of cluster orbitals on their character in the two former rings (let us refer to them as top and bottom ones). Figure 6 shows three examples; and the orbitals are given in two views for clarity. The first one with an energy of -0.738 a.u. is of the definitely σ kind and looks similarly in both views. Another one with an energy of -0.711 a.u. is of the clearly π kind in the top and bottom rings and of the prevailing bonding σ character within the side rings. What is important here are the relative and absolute energies of the orbitals. Describing a π electron density distribution in the basic molecular rings, the latter orbital is energetically very close to the boding σ orbital with an energy difference no larger than the one between two lower-energy σ cluster orbitals of two different kinds in the individual tetramer. Furthermore, now the delocalization of the orbital over the whole homodromic top and bottom rings of the structure is absolutely clear due to the higher density within the H-bonding regions.

This result can be interpreted as a kind of a tight stacking interaction between the top and bottom tetramolecular rings. This is a very distinctive feature of the cluster observed because of its overall symmetry.

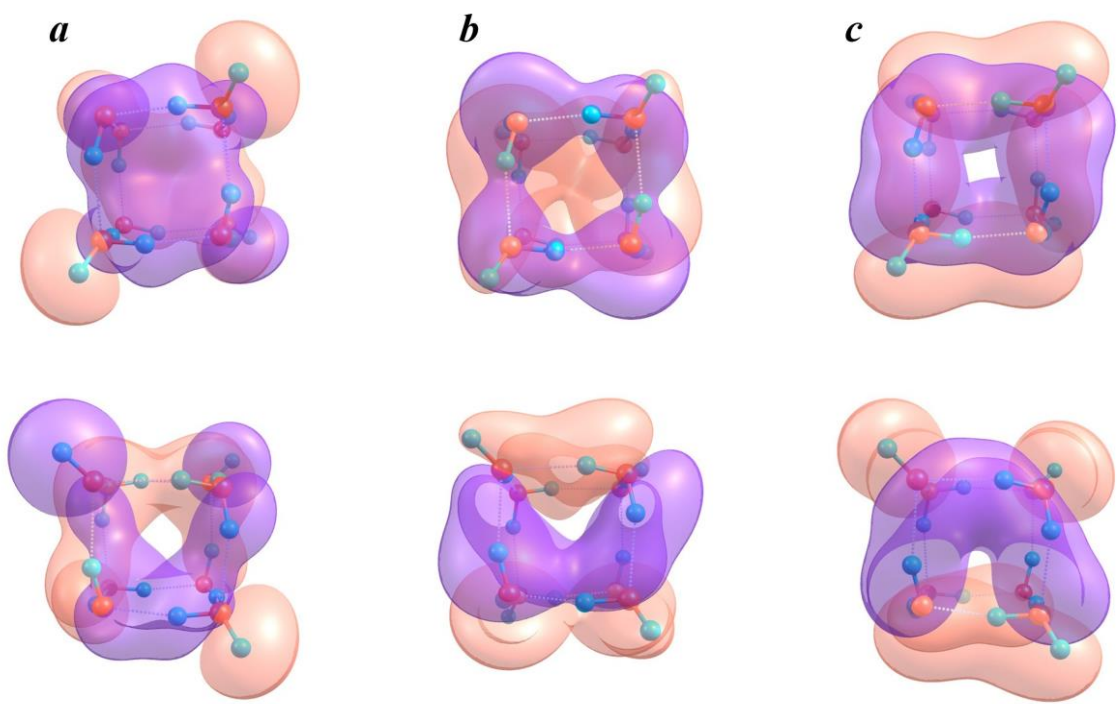

**Figure 6.** Cluster orbitals of $(H_2O)_8$ cubic octamer shown in two views (top) homodromic tetramolecular ring and (bottom) side non-homodromic tetramolecular ring (the orbitals are shown as transparent to make their spatial character clearer).

In the no less regular and symmetric dodecamer (Figure 1c), despite the large number of atoms and basis functions used for the approximation of the cluster orbitals, it is still possible to distinguish orbitals with clearly prevailing contributions of the oxygen atomic functions ($2p_x$, $2p_z$, and $2p_y$) and accordingly the prevailing σ or π kind of their overlapping. Some of those of the predominantly bonding character are shown in Figure 7. Their energies are -0.735, -0.7156, and -0.706 a.u. It is interesting that the first of these orbitals (Figure 7a) is of the bonding σ kind and envelops the whole cluster in such a way that its sign is constant within the side homodromic rings and alternating within the neighboring H-bond segments of the hexamolecular rings. The next one (Figure 7b) is again bonding within all H-bond domains, but now it looks like a skewed coat of the molecular cylinder composed of alternated sign segments, which are (conditionally) positive in the domains of those six molecules whose OH bonds are most elongated and negative in the domains of the residual again similar molecules. The third orbital is of bonding π kind within the top and bottom hexamolecular rings and of bonding σ kind in the H-bond domains of the side tetramolecular rings. And the energies of all the orbitals fall in the narrowest range of all those discussed above, which can be treated as a manifestation of the strongest coupling between the two kinds of bonding effects. Thus, here the mixing of π and σ characters is seemingly the most pronounced; and the role of delocalized π binding is the largest.

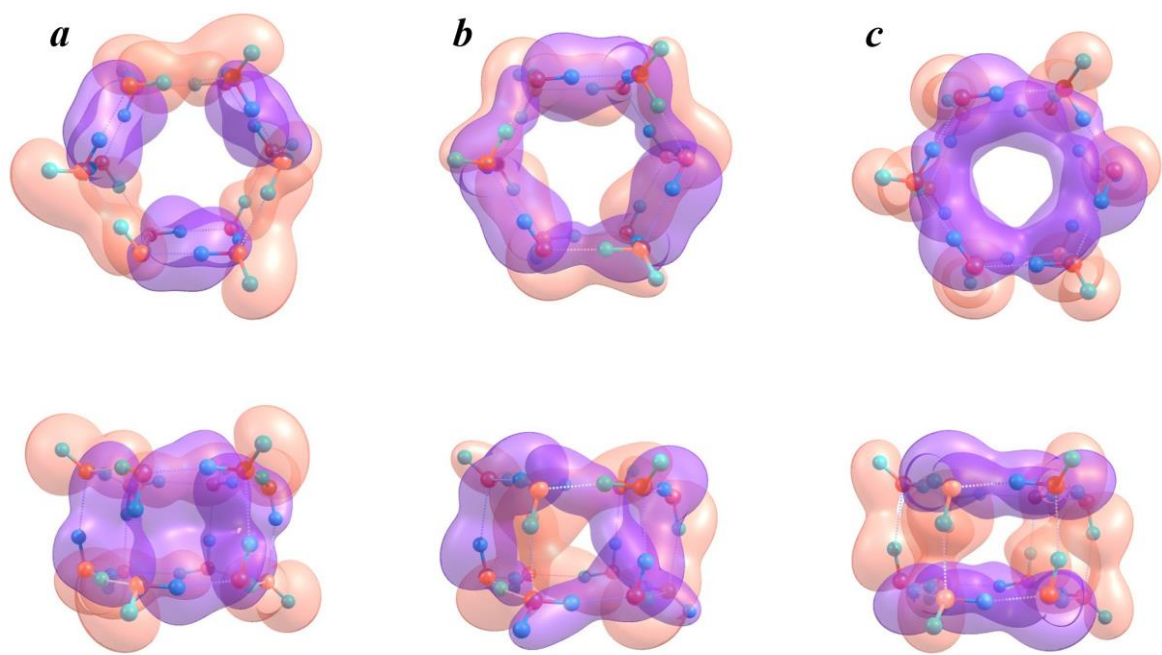

**Figure 7.** Cluster orbitals of $(H_2O)_{12}$ regular dodecamer shown in two views (top) homodromic hexamolecular ring and (bottom) side homodromic tetramolecular ring (the orbitals are shown as transparent to make their spatial character clearer).

However, all the features distinguished are no more than approximate characters of individual orbitals; whereas the gross (especially collective) effects of H-bond networks should be predetermined by the total electron density distribution, to which not only bonding, but also nonbonding and antibonding cluster orbitals contribute. To analyze the electron density distribution, we consider its response to the external magnetic field and quantify the interactions in terms of the charge redistribution that accompanies the formation of hydrogen bonds when these are expected to be coupled according to the above analysis of cluster orbitals.

*Atomic shielding in water clusters*

If we turn to the quantitative characteristics of the electron density distribution in the domains of hydrogen bonds, it is expedient to base the discussion on the comparison of the atomic shielding tensors and their eigenvectors with a reference to those in an individual water molecule. In the latter, these are as follows (in ppm) at the simulation level selected:

$$\sigma(O^1) = \begin{pmatrix} 378.6 & 0.0 & 0.0 \\ 0.0 & 351.0 & 0.0 \\ 0.0 & 0.0 & 332.1 \end{pmatrix}; \ \sigma(H^2) = \sigma(H^3) = \begin{pmatrix} 44.2 & 0.0 & 0.0 \\ 0.0 & 24.9 & 0.0 \\ 0.0 & 0.0 & 23.4 \end{pmatrix}$$

Note that the eigenvectors of the shielding tensor of oxygen nearly coincide with the principal axes of the molecule, the second being the $C_2$ axis; the first, a direction normal to it in the molecular plane; and the third, the direction of the oxygen lone-pair localization (Figure 8). The condensed parameters of the atomic shielding tensors are as follows:

| Atom | $\sigma_{iso}$, ppm | $\Delta\sigma$, ppm |
|---|---|---|
| $O^1$ | 353.9 | 37.1 |
| $H^2$, $H^3$ | 30.8 | 20.1 |

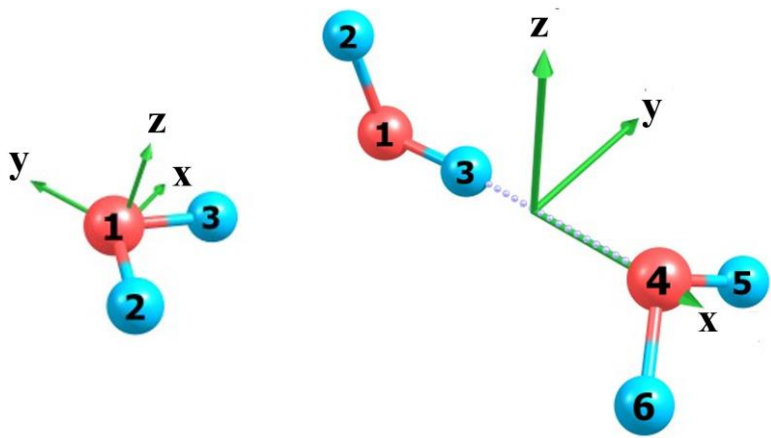

**Figure 8.** Water monomer and dimer with the numeration of atoms and the directions of principal axes shown.

As can be seen, the anisotropy of the shielding tensors of hydrogen atoms is mainly due to the existence of covalent bonds: the highest electron density is observed within the molecular plane (with a clear correlation with the contributions to O–H bonding molecular orbitals) as can be judged from the tensor represented in the principal-axes frame of reference:

$$\sigma(H^2) = \begin{pmatrix} 38.9 & 9.7 & 0.0 \\ 7.6 & 30.1 & 0.0 \\ 0.0 & 0.0 & 23.4 \end{pmatrix}$$

In the case of an oxygen atom, the largest shieldings are again typical of the same directions of covalent bonds, while the smallest (though still noticeable), in the direction of the lone pair localization.

What happens when two molecules become joined via a hydrogen bond, one of them being a proton donor, while the other, a proton acceptor (Figure 8)? The atomic shielding tensors of the water dimer in the frame of reference shown in Figure 8 are as follows:

$$\sigma(O^1)=\begin{pmatrix} 376.3 & 0.0 & -10.4 \\ 0.0 & 325.2 & 0.1 \\ -18.0 & 0.1 & 349.3 \end{pmatrix}; \sigma(O^4)=\begin{pmatrix} 343.2 & 0.0 & -7.6 \\ 0.0 & 366.8 & -0.1 \\ -8.6 & -0.1 & 330.7 \end{pmatrix}$$

$$\sigma(H^2)=\begin{pmatrix} 30.1 & 0.0 & -8.9 \\ 0.0 & 23.6 & 0.0 \\ -7.5 & 0.0 & 39.8 \end{pmatrix}; \sigma(H^3)=\begin{pmatrix} 48.7 & 0.0 & 0.9 \\ 0.0 & 16.7 & 0.0 \\ -0.8 & 0.0 & 18.1 \end{pmatrix}$$

$$\sigma(H^5)=\begin{pmatrix} 27.4 & 6.0 & -3.5 \\ 7.4 & 37.5 & -6.7 \\ -3.5 & -5.4 & 25.6 \end{pmatrix}; \sigma(H^6)=\begin{pmatrix} 27.4 & -6.0 & -3.5 \\ -7.4 & 37.5 & 6.7 \\ -3.5 & 5.4 & 25.6 \end{pmatrix}$$

In the direction that coincides with the OO line and is close to the $O^1$–$H^3$ covalent bond, the shielding of the oxygen atom of the proton donating molecule remains nearly unchanged while that of the proton wighdrawing one noticeably decreases, which means that this molecule provides its electron density to the formation of σ-kind binding within the H-bond domain. The increase in the $\sigma_{xx}$ element of $H^3$ atom to 48.7 ppm shows that the electron density in the direction of H-bonding within $O^1$–$H^3$…$O^4$ fragment actually increases, which is an additional indication of the intermolecualr σ-binding.

If we turn to the diagonalized shielding tensors listed below and analyze their eigenvectors, it can be shown that projections of the first eigenvector of the σ($H^3$) tensor on the $O^1$…$O^4$ and $H^3$…$O^4$ lines equal nearly unity: 0.998 and 0.999, respectively. This means that the quite noticeable charge polarization is governed by the O…O interaction mediated by the bridge proton.

$$\sigma(O^1)=\begin{pmatrix} 382.2 & 0.0 & 0.0 \\ 0.0 & 325.2 & 0.0 \\ 0.0 & 0.0 & 343.5 \end{pmatrix}; \sigma(O^4)=\begin{pmatrix} 366.8 & 0.0 & 0.0 \\ 0.0 & 347.1 & 0.0 \\ 0.0 & 0.0 & 326.7 \end{pmatrix}$$

$$\sigma(H^2)=\begin{pmatrix} 44.4 & 0.0 & 0.0 \\ 0.0 & 25.5 & 0.0 \\ 0.0 & 0.0 & 23.6 \end{pmatrix}; \sigma(H^3)=\begin{pmatrix} 48.7 & 0.0 & 0.0 \\ 0.0 & 18.1 & 0.0 \\ 0.0 & 0.0 & 16.7 \end{pmatrix}$$

$$\sigma(H^5)= \sigma(H^6)=\begin{pmatrix} 43.5 & 0.0 & 0.0 \\ 0.0 & 24.3 & 0.0 \\ 0.0 & 0.0 & 22.7 \end{pmatrix}$$

It is worth noting that condensed parameters of the $\sigma(O^1)$ and $\sigma(O^4)$ tensors can provide only partial information. For example, the isotropic shieldings of both atoms slightly decrease (to 350.2 and 346.9 ppm respectively), which means that a certain amount of the electronic charge is redistributed to provide the intermolecular binding, but one cannot guess which atom acts as a charge donor. Anisotropy of the tensors give certain ideas about the redistribution, but again relatively general: it increases in the case of $O^1$ to 48.2 ppm and decreases in the case of $O^4$ to 29.9 ppm. Taking into account that the anisotropy of shielding of an oxygen atom in an individual water molecule is chiefly due to the difference between the electron density in the domains of covalent bonds and the lone-pair density, the latter decrease can be due to the partial donation of the charge density to the intermolecular contact. This is accompanied by the substantial increase in the anisotropy of the charge distribution around $H^3$ proton: it reaches 31.3 ppm at a mean isotropic value quite close to those of the residual hydrogen atoms, about 27.8 ppm. This result also supports the conclusion about the predominant σ-kind directional charge redistribution within the $O^1$-$H^3$…$O^4$ H-bond domain.

Let us turn now to larger clusters where the coupling between the individual molecular orbitals of not only σ, but also π kind was distinguished (see previous section). In a homodromic water tetramer, the condensed shielding characteristics of all the atoms of the same kind (O oxygen atoms, $H_b$ bridge hydrogen atoms and $H_d$ dangling hydrogen atoms uninvolved in hydrogen bonds) are identical due to the symmetrical structure with absolutely equivalent neighborhoods of all the molecules, each of which acts as a single proton donor and a single proton acceptor in the contacts with the close neighbors:

| Atom | $\sigma_{iso}$, ppm | $\Delta\sigma$, ppm |
|---|---|---|
| O | 335.1 | 41.1 |
| $H_b$ | 24.9 | 33.3 |
| $H_d$ | 30.4 | 20.3 |

As can be seen, the anisotropy of shielding of both oxygen and hydrogen atoms involved in H-bonding increases, while the isotropic shielding of both kinds of atoms decreases. This general result can be interpreted only as a donation of a certain fraction of electronic charge to the H-bond domains. At the same time, the anisotropy of shielding of the bridge protons is even larger than in a water dimer at a smaller isotropic value, which means that the electron density around bridge protons in the tetrameric ring becomes more polarized compared to an individual H-bond. The diagonalized shielding tensors of the unique atoms in the cluster enable one to make the conclusions more accurate:

$$\sigma(O) = \begin{pmatrix} 362.2 & 0.0 & 0.0 \\ 0.0 & 332.1 & 0.0 \\ 0.0 & 0.0 & 311.0 \end{pmatrix};$$

$$\sigma(H_d) = \begin{pmatrix} 44.0 & 0.0 & 0.0 \\ 0.0 & 24.3 & 0.0 \\ 0.0 & 0.0 & 23.0 \end{pmatrix}; \quad \sigma(H_b) = \begin{pmatrix} 47.1 & 0.0 & 0.0 \\ 0.0 & 15.6 & 0.0 \\ 0.0 & 0.0 & 12.1 \end{pmatrix}$$

As can be noticed, the decrease in the eigenvalues of the shielding tensors of oxygen atoms is nearly uniform, which means that the decrease (or donation) of the charge density takes place not only in the directions of H-bonds (two mutually orthogonal directions in a mean oxygen plane of the structure), but also in the normal direction. This means that the pre-lone pair density is spent on the formation of additional electronic bonding, which is actually π bonding. The first eigenvector of any $\sigma(H_b)$ tensor nearly accurately (with a projection of 0.997) coincides with the corresponding O…O line, and the eigenvalue is higher than that of $\sigma(H_d)$ for the dangling hydrogen atoms. At the same time, the deshielding of bridge protons in the two normal directions is very pronounced. And if the decrease in the second eigenvalue can be caused by the electron density redistribution within the molecular plane due to the directional σ contribution to the H-bonding, the decrease in the third value can be atributed only to the location of the bridge protons close to the minima of the π-bonding cluster orbitals and the zero points of the π-antibonding orbitals.

To check the conclusions let us turn to the situation when two homodromic tetrameric rings are joined via a common molecule with a resulting *ddaa* coordination (Figure 9a). Because of the approximate orthogonality of the mean molecular planes of the two tetrameric subunits, their mutual effect is not as pronounced (see the above discussion of the molecular orbitals). In fact, the mean condensed shielding characteristics of the atoms of all the residual molecules (with *da* coordination) are quite close to those of an individual tetramer; and only those of the central molecule, which is involved in four hydrogen bonds, differ:

| Atom (molecule) | $\sigma_{iso}$, ppm | $\Delta\sigma$, ppm |
|---|---|---|
| O (*da*) | 334.2 | 42.6 |
| H$_b$ (*da*) | 24.9 | 33.8 |
| H$_{nb}$ (*da*) | 30.4 | 20.3 |
| O (*ddaa*) | 313.0 | 36.7 |
| H$_b$ (*ddaa*) | 24.6 | 33.2 |

The strongest difference is observed in the case of the $O^{13}$ oxygen atom:

$$\sigma(O^{13}) = \begin{pmatrix} 337.2 & 0.0 & 0.0 \\ 0.0 & 302.2 & 0.0 \\ 0.0 & 0.0 & 299.5 \end{pmatrix}.$$

At the smaller anisotropy, which reflects the actual closeness of the electron density distribution around the O nucleus to more or less distorted tetrahedral, the isotropic value is smaller by more than 20 ppm compared to the *da* molecules. A similar absolute decrease characterizes the involvement of a water molecule in two H-bonds (compare water monomer and tetramer). What is interesting, the eigenvectors of this shielding tensor do not coincide with the directions of the O…O lines between the central molecule and its closest neighbors. The projections of the first vector on $O^{13}…O^{1}$ and $O^{13}…O^{19}$ vector lines are $\pm 0.85$ and those of the second vector on $O^{13}…O^{4}$ and $O^{13}…O^{10}$ lines equal $\pm 0.75$, but projections of the vectors on $O^{1}…O^{19}$ and $O^{4}…O^{10}$ lines are nearly unit. This result shows that the polarization of the electron density distribution is driven not only by the local neighbors as was in a tetramer, but rather by the more extended correlation between the H-bonded particles. Here, a clear cross correlation within the central pentamolecular fragment of the cluster is present, which shows that the two ring subunits are actually coupled via the central molecule. Note that the eigenvectors of the shielding tensors of all bridge hydrogen atoms (which correspond to the largest eigenvalues) practically coincide with the O…O directions of the respective hydrogen bonds.

Thus, oxygens of the directly H-bonded water molecules govern the electron density redistribution around the bridge hydrogen nucleus. At the next organization level, again the oxygens of the two molecules that are located at opposite sides of the third (middle) molecule govern the charge redistribution around its oxygen nucleus irrespectively of whether the former belong to the same or different ring structure elements of the H-bond network. This is a direct indication of the coupling between the hydrogen bonds not only within the same homodromic ring, but also of the rings joined via a common tetracoordinate molecule.

Here, the *da* molecules that act as electron density acceptors in their H-bonds with the central molecule are characterized by the largest anisotropy of shielding, about 43.9 ppm; whereas those that act as electron density donors to the central molecule, by the smallest anisotropy of 40.4 ppm. Although the values differ not as substantially, effects produced by the relative electron density deficiency can be noticed. Nevertheless, on the whole the shielding tensors of all the *da* molecules, nearly irrespectively of whether these are close neighbors of the central one or separated from it, have quite similar eigenvalues in reasonable closeness to those of a ring-like tetramer:

$$\sigma(O^1) = \begin{pmatrix} 361.5 & 0 & 0 \\ 0 & 333.5 & 0 \\ 0 & 0 & 309.8 \end{pmatrix} \text{ vs. } \sigma(O^7) = \begin{pmatrix} 362.6 & 0 & 0 \\ 0 & 331.3 & 0 \\ 0 & 0 & 311.5 \end{pmatrix}$$

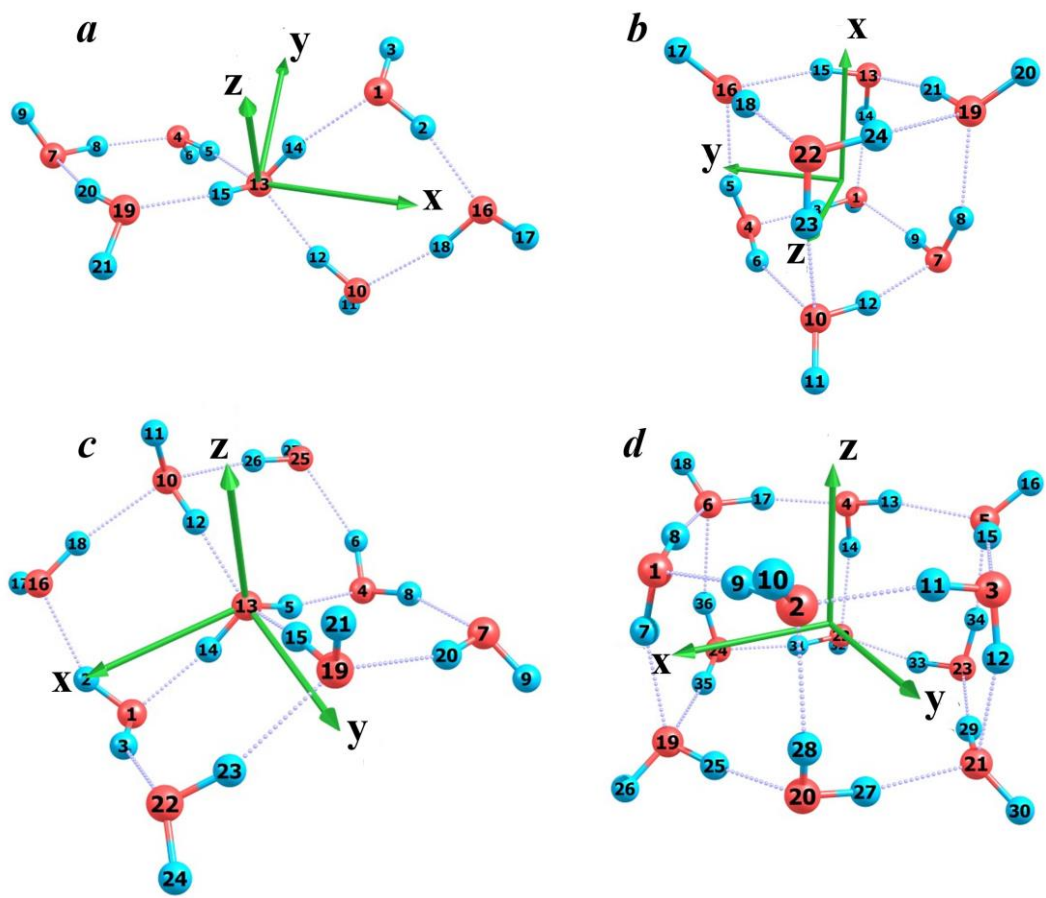

**Figure 9.** Water clusters with the numeration of atoms and the directions of principal axes shown: (a) dicyclic heptamer, (b) cubic octamer; (c) nonamer composed of four fused homodromic tetramers; and (d) dodecamer, the structure of which involves top and bottom hexamolecular homodromic rings bound via tetramolecular side faces.

What happens when the number of close neighbors of most of the water molecules in a cluster increases to three, which is already typical of condensed phase specimens? Let us first consider a cubic octamer, which can be viewed as composed of two homodromic tetramers bound to each other (Figure 9b). A half of the molecules in it have *dda* local coordination, while the other half are of *daa* kind. It is reasonable to look at first at the shielding tensors of some unique atoms in the cluster in the principal-axes frame of reference:

$$\sigma(O^{10}) = \begin{pmatrix} 319.3 & -3.0 & -16.7 \\ 0.2 & 323.5 & 14.9 \\ -13.9 & 18.6 & 339.1 \end{pmatrix}; \sigma(O^4) = \begin{pmatrix} 329.1 & 12.4 & -6.3 \\ 11.5 & 318.2 & -4.6 \\ -5.1 & -4.4 & 324.1 \end{pmatrix};$$

$$\sigma(H^{12})=\begin{pmatrix} 12.2 & -0.9 & -3.7 \\ -0.3 & 28.5 & 16.7 \\ -3.0 & 17.7 & 28.9 \end{pmatrix}; \sigma(H^5)=\begin{pmatrix} 43.3 & 5.4 & 3.4 \\ 5.2 & 19.9 & 0.3 \\ 2.4 & 0.1 & 17.2 \end{pmatrix}.$$

As can be seen, z-direction in the case of $H^5$ atom and x direction in the case of $H^{12}$ atom are quite close to the eigenvectors of the tensors. Despite the fact that the directions in both cases are orthogonal to the corresponding H-bonds, the diagonal elements of the tensors are noticerably different; and the deshielding of $H^{12}$ atom is much more pronounced. Note that $H^{12}$ atom belongs to a homodromic ring where π contribution to the intermolecular bonding is more substantial (see above), while $H^5$ atom is located within that very π-kind sandwich of this cibic cluster. Then, the difference between the shielding parameters agrees quite reasonably with the idea about combined π/σ character of the H-bonding.

To compare the shielding characteristics of oxygen atoms, let us look at the diagonalized tensors and their condensed characteristics:

$$\sigma(O^{10})=\begin{pmatrix} 354.9 & 0.0 & 0.0 \\ 0.0 & 320.2 & 0.0 \\ 0.0 & 0.0 & 306.8 \end{pmatrix}; \sigma(O^4)=\begin{pmatrix} 340.0 & 0.0 & 0.0 \\ 0.0 & 320.9 & 0.0 \\ 0.0 & 0.0 & 310.4 \end{pmatrix}$$

$$\sigma(H^{12})=\begin{pmatrix} 46.1 & 0.0 & 0.0 \\ 0.0 & 13.7 & 0.0 \\ 0.0 & 0.0 & 9.8 \end{pmatrix}; \sigma(H^5)=\begin{pmatrix} 44.8 & 0.0 & 0.0 \\ 0.0 & 18.9 & 0.0 \\ 0.0 & 0.0 & 16.9 \end{pmatrix}$$

| Atom (molecule) | $\sigma_{iso}$, ppm | $\Delta\sigma$, ppm |
|---|---|---|
| $O^{10}$ (*daa*) | 327.3 | 41.7 |
| $O^4$ (*dda*) | 323.8 | 24.4 |
| $H^{12}$ (*daa*) | 23.2 | 34.4 |
| $H^5$ (*dda*) | 26.8 | 26.9 |

The isotropic shieldings of the oxygen atoms involved in *daa* and *dda* molecules ($O^{10}$ and $O^4$ respectively) are very close, whereas the anisotropy drastically differs, being very small in the *dda* molecule. Note that this molecule is involved in three H-bonding interactions in nearly mutually orthogonal directions, two of which practically coincide with covalent bonds and the third, with the hydrogen bond. The lowest shielding anisotropy of the oxygen atom reflects such a redistribution of the electron density that minimizes (definitely not vanishing) the difference between the three directions, in two of which the oxygen acts as an electron acceptor (being a proton donor) and in one, as an electron donor (being a proton acceptor).

In the case of hydrogen atoms, the one that belongs to a homodromic ring ($H^{12}$) is characterized by an extremely low shielding in a direction normal to its covalent ($O^{10}$–$H^{12}$) bond, where the shielding is the largest. If referenced to an individual tetramer, this can be explained by a substantial electron density redistribution toward another (neighboring) homodromic ring.

Now let us turn to a nonamer composed of four fused homodromic tetramolecular rings, in which the central molecule finds itself in an improper local geometry because of the general convex-concaved configuration (Figure 9c). If compared to the heptamer considered above, two additional molecules act here as ties between the two rings, which makes the mutually orthogonal arrangement of the rings impossible. The angle between the planes of tetrameric rings in the heptamer was close to 90°, whereas here, the angles between the planes of similarly arranged tetrameric subunits are close to 60°.

The latter peculiarity results in the so far lowest isotropic shielding (294.5 ppm) of the central molecule at a relatively low anisotropy (35.6 ppm):

$$\sigma(O^{13}) = \begin{pmatrix} 318.3 & 0.0 & 0.0 \\ 0.0 & 291.8 & 0.0 \\ 0.0 & 0.0 & 273.5 \end{pmatrix}$$

Compared to the tetracoordinate molecule in the heptamer, all the eigenvalues of the tensor are lower by 10 to 25 ppm. This is seemingly predetmined by the impossibility to compensate the charge density donated to the formation of some H-bonds (in which the molecule acts as a proton acceptor) by efficiently withdrawing the density from those molecules in the H-bonds with which it acts as a proton donor. As to the molecules that are close neighbors of the central one, half of them are of *dda* kind, the residual, of *daa* kind; and their condensed shielding characteristics are as follows:

| Atom (molecule) | $\sigma_{iso}$, ppm | $\Delta\sigma$, ppm |
|---|---|---|
| O (*daa*) | 323.4 | 41.4 |
| O (*dda*) | 324.8 | 40.4 |
| H (*daa*) | 22.7 | 37.3 |
| H (*dda*) | 26.8 | 28.1 |

As can be seen, the parameters of bridge hydrogen atoms are nearly the same as in the molecules of the same coordination kinds in the cubic octamer. The same can be said about the oxygen atom in *daa* molecule; and nearly the same values characterize the atom in *dda* molecule, which makes it drastically different from the atoms in similar neighborhoods in the octamer, where the anisotropies of shielding were much smaller. The high anisotropy reflects the substantial charge polarization, which can readily be judged from the diagonalized shielding tensors:

$$\sigma(\text{O(in } dda))= \begin{pmatrix} 351.8 & 0.0 & 0.0 \\ 0.0 & 321.4 & 0.0 \\ 0.0 & 0.0 & 301.3 \end{pmatrix}; \quad \sigma(\text{O (in } daa))= \begin{pmatrix} 350.6 & 0.0 & 0.0 \\ 0.0 & 312.7 & 0.0 \\ 0.0 & 0.0 & 306.9 \end{pmatrix}$$

If we additionally take a look at the shielding tensors of all the four bridge hydrogen atoms in the central part of the structure

$$\sigma(H^5) \approx \sigma(H^{14}) = \begin{pmatrix} 46.3 & 0.0 & 0.0 \\ 0.0 & 11.8 & 0.0 \\ 0.0 & 0.0 & 9.0 \end{pmatrix}; \quad \sigma(H^{12}) \approx \sigma(H^{15}) = \begin{pmatrix} 47.6 & 0.0 & 0.0 \\ 0.0 & 12.0 & 0.0 \\ 0.0 & 0.0 & 8.6 \end{pmatrix},$$

we see that the deshielding in z direction is typical of all oxygen atoms (except for $O^{13}$) and all bridge hydrogen atoms which are involved in two homodromic rings each; and $O^{13}$ atom is involved in as many as four homodromic rings simultaneously. The latter seems nearly impossible in any real sample for the dynamic effects will prevent the formation of such a symmetric structure with an unfavorable steric neighborhood of the central molecule. However, as a theoretical reference case, the configuration is spectacular. It shows most clearly that the larger the coupling between the molecules (as provided by alternated hydrogen and covalent bonds) involved in homodromic rings, the stronger the electron density redistribution and polarization. Particularly,

$$\sigma_{iso}(O, ddaa) < \sigma_{iso}(O, dda, daa) < \sigma_{iso}(O, da)$$
$$\sigma_3(O, ddaa) < \sigma_3(O, dda, daa) < \sigma_3(O, da),$$

where $\sigma_3$ is the lowest eigenvalue of the shielding tensor.

To complete the discussion, let us consider the structure where all molecules are tricoordinate, a half of them being involved in one homodromic ring, and the residual, in two rings; and the rings are of different molecular size. This is the regular dodecamer configuration (Figure 9d), in which two (top and bottom) homodromic hexamolecular rings are bound via side tetramolecular rings, every other of which is homodromic. In the structure, all oxygen atoms are involved in two homodromic rings, but their local coordination neighborhoods differ, being of either *dda* or *daa* kind. At the same time, bridge hydrogen atoms are of two kinds depending on the number of homodromic rings they are involved in.

In a *daa* molecule, the shielding of an oxygen atom is larger in the direction of its O…H–O contacts within the hexamolecular ring compared to the tetramolecular one (by ca. 17 ppm); being the largest in the direction of electron withdrawing interaction within the hexamolecular ring. Similarly, in a *dda* molecule, the largest deshielding of the oxygen atom (which is electron withdrawing in two O–H…O bonds) is observed in a direction normal to the homodromic hexamolecular rings (σ is about 305 ppm) and the smallest, in two residual directions (σ is about 322 and 348 ppm), which can be interpreted as a stronger π-coupling between the H-bonded molecules within a larger ring where the local inclination angles of the molecules provide better conditions for the π-overlapping. On the whole, at the close isotropic shieldings of the atoms, the anisotropy is higher in the case of *daa* molecules: 36.0 vs. 29.0 ppm.

As to the hydrogen atoms, those that act as bridges in two homodromic rings simultaneously, are characterized by the so far strongest polarization of the electron density distribution:

$$\sigma(H^{17}) = \begin{pmatrix} 47.0 & 0.0 & 0.0 \\ 0.0 & 12.2 & 0.0 \\ 0.0 & 0.0 & 8.8 \end{pmatrix};$$

so that the averaged condensed shielding parameters are as follows: $\sigma_{iso}$= 22.7 ppm and $\Delta\sigma$ = 36.6 ppm. Here, the largest shielding is observed in the direction of the H-bond common for the two rings; while the very low value characterizes the charge decrease in the direction normal to the two H-bonded rings. As can be seen, the condensed parameters can give no idea about the actual charge distribution if the shielding tensor is unknown.

In *dda* molecules, one bridge hydrogen atom is involved in a hexamolecualr ring, while the other in a tetramolecular one. Their formal condensed shielding parameters are not as different:

$\sigma_{iso}$= 26.9 ppm and $\Delta\sigma$ = 26.6 ppm within a tetramolecular ring;

$\sigma_{iso}$= 26.0 ppm and $\Delta\sigma$ = 30.7 ppm within a hexamolelecular ring,

although a slightly larger density in the H-bond direction and a slightly smaller one in the normal directions can be noticed in the case of that involved in a hexamolecular ring as reflected in a little bit larger anisotropy.

**CONCLUSIONS**

The electron density distributions in the vicinities of both oxygen and bridge hydrogen nuclei in H-bonded water networks are very sensitive to the local coordination neighborhoods of the molecules. At the same time, the isotropic shielding, which can be retrieved from experimental data, is not as informative for the drastically different polarization of the charge distributions depending on the effects produced by their close neighbors.

Nevertheless, even the isotropic shieldings of oxygen nuclei are informative when considered for the molecules with different neighborhoods. Based on the above data, formal chemical shifts with respect to (gas-phase) monomer are as follows: ca. -19…-20 ppm in the *da* molecules; -27…--30 ppm in *daa* molecules; -29…-31 ppm in *dda* molecules; and ca. -41…-59 ppm in *ddaa* molecules. As was already mentioned, there is no substantial difference between the tricoordinate molecules irrespectively of whether they act as double proton donors or double acceptors, since they are involved in three coupling interactions with the neighbors in the directions not as strongly deviating from mutual orthogonality; although a larger anisotropy is observed in the case of $^{17}$O shielding in *daa* molecules. At the same time, the appearance of four close neighbors gives rise to a substantial increase in deshielding, which becomes more pronounced with an increase in local steric hindrances that hamper reaching a certain counterbalance between the electron-withdrawal and donation in the formation of hydrogen bonds. Thus, the mean value of -36 ppm accepted as the one characteristic of liquid water is an average estimate, which should change with the change in the fractions of molecules with different neighborhoods. This is what particularly takes place with an increase in temperature when the fraction of tetracoordinate molecules decreases, and, judging from the quoted value of -26 ppm at 215°C, nearly vanishes under such conditions.

However, these data provide no idea about the character of electronic binding in the case of hydrogen bonds. At the same time, principal values of the shielding tensors reflect the degree and direction of the electron density redistribution that accompanies the bond formation. The shielding of dangling protons uninvolved in hydrogen bonds remains at an almost the same level as in an individual water molecule. The shielding of the bridge proton noticeably increases in the direction that coincicdes with the O…O line of the bond, which shows that it is the σ-overlapping of p-type oxygen orbitals that governs the directionality of hydrogen bonds; and the mediating bridge hydrogen atom is accurately built in thus appeared intermolecular contact.

Moreover, a second level of correlation can be distinguished when the charge density in pentamolecular fragments with the central *ddaa* molecule is analyzed. The eigenvectors of the shielding tensor of the central molecule nearly coincide with the practically orthogonal O…O lines formed by the pairs of oxygen nuclei in its close neighbors opposing each other in the local tetrahedron. Thus, here already the tetracoordinate molecule acts as a mediating element in the so-to-speak 1–3 coupling of molecules in the hydrogen bond network.

Finally, the nature of coupling can be clarified based on the principal values and eigenvectors of the shielding tensors of oxygen nuclei. As clearly follows from the above discussion, the deshielding is the largest in the direction normal to a homodromic ring. The deshielding here is caused by the electron density redistribution related to the π-overlapping of pre-lone pairs of oxygen atoms within the H-bond domains. Such deshielding is typical of any kinds of double bonds in the compounds with conventional π-binding. In the case of hydrogen bonds, the correspodning principal values of the shielding tensors of O nuclei differ from the largest ones by up to 50 ppm, which shows that actually the contribution of this kind of binding is substantial even taking into account the nearly twice as large internuclear distances (compared to typical intramolecular bonds).

Thus, the directionality of hydrogen bonds is predetermined by σ-binding, while the coupling and collective effects within H-bond networks, by π-binding between water molecules, in both cases the overlapping of p-type orbitals of oxygen atoms governing the interaction at the mediating role of bridge protons.

**ACKNOWLEDGMENTS**

The authors would like to acknowledge the help of Dr. D.A. Bokhan in carrying out quantum chemical simulations with the use of licensed CFOUR software.